\documentstyle[12pt,epsfig]{article}

\begin{document}

\begin{center}
{\large \bf 
On Relation Between the Quark Gluon Bag Surface Tension\\ 
and the Colour Tube String Tension}

%%\shorttitle{Surface Tension of QG Bag and the Colour Tube String Tension} 

\vspace{1.0cm}

{\bf K. A. Bugaev  and G. M. Zinovjev}

\vspace{1.cm}

Bogolyubov Institute for Theoretical Physics,
National Academy of Sciences of Ukraine, \\
Metrologichna str. 14$^B$,  03680 -- Kiev, Ukraine\\

\vspace{1.cm}

\end{center}

\abstract{
{Here we revisit the bag phenomenology of the deconfining phase transition
to replenish it by introducing systematically the bag surface tension. 
Comparing the free energy of such bags and that one of  the strings confining the static quark-antiquark pair, 
we express the string tension in terms of the bag surface tension and
thermal pressure in order to estimate the bag characteristics using the lattice QCD data. 
Our analysis of the bag entropy density demonstrates that the surface tension coefficient is   
amazingly negative at the cross-over (continuous transition). This approach allows us to naturally account 
for an appearance of a very pronounced maximum (observed in the lattice QCD simulations)
of the entropy of the bound static quark-antiquark pair. The vicinity of the (tri)critical 
endpoint  is also  analyzed to clarify the meaning  of vanishing surface tension coefficient.} 
\\

\noindent
{\bf Key words:}  Surface free energy, surface tension, string tension, string radius\\
{\bf PACS:} 25.75.Nq, 25.75.-q
}

%%{25.75.Nq = Quark deconfinement, quark gluon plasma production, and phase transitions}
%%{25.75.-q = Relativistic heavy-ion collisions}

%%%%%%     03.05.2010          Kyrill             NPA_1.tex

\section{Introduction}
One of the key physical quantities provided by the lattice quantum chromodynamics (LQCD) is the 
free energy of static quark-antiquark pair $F_{q\bar q} (T, L)$ as a function of the temperature
$T$ and separation distance $L$ being extracted from the Polyakov line correlation in a colour 
singlet channel. The linear  $L$-dependence of $F_{q\bar q} (T, L)$ discovered at large $L$ and 
low temperatures naturally explains the colour confinement. On the other hand at higher 
temperatures the linear $L$-dependence of $F_{q\bar q} (T, L)$ disappears {signaling 
the Debye screening and an advent of deconfined phase. This phenomenologically transparent 
picture resembles the confining string model  \cite{Patel} which is fully adopted by the LQCD 
community \cite{ConfString1,ConfString2,ConfString3}.}

{Here we develop another general 
%%%complementary 
view of the confinement phenomenon dealing entirely with an idea of quark-gluon (QG) bag with 
nonzero surface tension. In this approach}
the importance of the  surface tension concept was realized long ago \cite{Jaffe:1,Svet:1}, but  
only recently the surface tension of large QG bags was consistently included into the statistical  
description of the QG plasma equation of state \cite{Bugaev:07, FWM, Reggeons:08, Bugaev:08,CritPoint}.
It turns out quite suitable
%%this not only allows one 
to formulate the analytically solvable statistical models for the QCD tricritical \cite{Bugaev:07} and 
critical \cite{CritPoint} endpoint, and to push forward an idea of the finite width model of QG bags 
\cite{FWM, Reggeons:08, Bugaev:08}. Clearly such a development improves our understanding of the QG 
plasma equation of state and brings it forward to a qualitatively new level of realism by establishing 
the Regge trajectories of heavy/large bags both in a vacuum and in a medium \cite{FWM,  Reggeons:08,
Bugaev:08} using the LQCD data. However, to make the approach quantitatively informative one 
needs to establish the value of the surface tension in the whole (maximally possible) range of  
temperature and baryonic chemical potential. Unfortunately, as for now the LQCD cannot provide us with 
such an information. To resolve this problem we are going here to ascertain the phenomenological  
relation between the string tension of {a tube confining the static quark-antiquark pair} and  
the surface tension of the QG  bags {in order to study the bag thermodynamics. We believe the 
concept of bag surface tension looks more adequate just at high temperatures. Another major 
 (and closely related) task of this study is an investigation of the (tri)critical endpoint 
vicinity of the QCD phase diagram to clarify the meaning of vanishing surface tension coefficient.} 

%%%%%%%%%%%%%%%%%%%%%%%%%%%%  Sect II

\section{Free Energy of Elongated Cylindrical Bag}
The free energy of large almost spherical QG bag can be cast as \cite{Bugaev:07, FWM, Reggeons:08, 
Bugaev:08, CritPoint}
\begin{eqnarray}\label{EqI}
& \hspace*{-0.4cm}F_{sp}(T,V) =  - p_v (T) V + 2\, \sigma_{surf} (T) S(V) + T \tau \ln\left[ \frac{V}{V_0} \right] .~
\end{eqnarray}
Here $p_v (T)$ is the thermal pressure inside a bag, $\sigma_{surf} (T)$ is the temperature dependent 
surface tension coefficient, $S(V) \sim V^{\frac{2}{3}}$ is the mean surface of the QG bag, while the 
last term on the right hand side of (\ref{EqI}) is the Fisher topological term \cite{Fisher:67} which 
is proportional to the Fisher exponent $\tau = const > 1 $ \cite{Bugaev:07, CritPoint}.
$V_0$ is a normalization constant with the dimension of volume. Note that this is the standard 
parameterization of the free energy of large physical clusters which is successfully used in the Fisher 
droplet model \cite{Fisher:67}, in the statistical multifragmentation model \cite{Bondorf:95, Bugaev:00}  
and in describing the free energy of large geometrical clusters of the 2- and 3-dimensional Ising model  
\cite{Ising:clust, Complement} and of percolation clusters \cite{Percolation}. Such a free energy   
parameterization turns out very efficient in studying the critical point of realistic gases \cite{Fisher:67, 
Elliott:06}. It was applied to many different systems with the different extents of success including a 
nuclear multifragmentation both in infinite \cite{Bondorf:95,Bugaev:00,Moretto,Reuter:01,Bugaev:05c} and  
in finite systems \cite{Bugaev:05c}, a nucleation of real fluids \cite{Dillmann} and the compressibility 
factor of real fluids \cite{Kiang}. 

In principle, besides  the bulk $(\sim V)$ and surface parts $(\sim S(V))$ the free energy (\ref{EqI}) 
could include the curvature part as well, which may be important for small hadronic bubbles \cite{Svet:1} 
or for cosmological phase transition study \cite{Ignat:1}. We stress, however, that as usual the critical 
properties of the statistical  models are defined by the infinite bag, therefore, including a curvature 
term of any sign in (\ref{EqI}) could affect the thermodynamic quantities of such models at (tri)critical  
endpoint only \cite{Bugaev:07,CritPoint} (see also below). If the curvature term is of a real importance 
for the cluster models discussed here, then it should also show itself at the (tri)critical points of many 
systems described by free energy of Eq.(\ref{EqI}) \cite{Fisher:67, Bondorf:95, Bugaev:00, 
Ising:clust, Complement, Percolation, Elliott:06,
Moretto, 
Reuter:01, Bugaev:05c, Dillmann, Kiang}, but this is not 
the case {(see an extended discussion in \cite{Bugaev:07})}. Keeping in mind this argument we 
omit the curvature part of bag free energy as well.

Using the thermodynamic identity  
\begin{equation}\label{EqII}
p_{sp} = - \left( \frac{\partial F}{\partial V}\right)_T = p_v(T) - 2\, \sigma_{surf} (T)\frac{\partial S(V)}{\partial V} - \frac{T \tau}{V}\,,
\end{equation}
one finds the pressure of the spherical bag from (\ref{EqI}). To calculate the free energy of bags of more 
complicated shapes one has to change the second term on the right hand side of Eq.(\ref{EqII}) to the general 
Laplace form of surface pressure
\begin{equation}\label{EqIII}
p_{gen} =  p_v(T) -  \sigma_{surf} (T) \left[ \frac{1}{R_1} +  \frac{1}{R_2}  \right] - \frac{T \tau}{V}\,,
\end{equation}
where $R_1$ and $R_2$ are the main curvature radii of the shape defined locally. Basing on Eq.(\ref{EqIII}) 
one can find the free energy of an arbitrary shaped bag.
%%%up to a temperature dependent  function which, however, 
%%%does not play any role since we are interested in changes of free energy. 
Then for the large elongated cylinder of the radius $R = R_1$ and the height $L\gg R$ one gets ($R_2 = \infty$) 
\begin{eqnarray}\label{EqIV}
&&\hspace*{-0.5cm}F_{cyl} (T, L, R) \equiv - \int dV\, p_{gen} = 
- p_v(T) \pi R^2 L + \sigma_{surf} (T) 2 \pi R L + T \tau \ln V + f(T) \,.
\end{eqnarray}
Here the volume independent function $f(T)$ is the integration constant. Comparing Eq.(\ref{EqIV}) with 
Eq.(\ref{EqI}) and applying the same steps to obtain the sphere free energy from Eq.(\ref{EqII}), we 
conclude that $f(T) = - T \tau \ln V_0$ for a cylinder just like for a sphere. 

Assuming now that the free energy of cylindrical bag (\ref{EqIV}) of the radius $R$ and the length $L \gg R$   
equals to the free energy of colour string of the same size $F_{str} \approx \sigma_{str} L$ that 
binds 
%%%bounds   
the static quark-antiquark pair one can find the desired relation {at vanishing baryonic densities:}
\begin{equation}\label{EqV}
\sigma_{str} (T) = \sigma_{surf} (T)\, 2 \pi R~ - ~p _v (T) \pi R^2 + \frac{T \tau}{L} \ln\left[
\frac{\pi R^2 L }{V_0} \right] \,.
\end{equation}
{In doing so we match an ensemble of all string shapes of fixed $L$ to a mean elongated cylinder, 
which according to the original Fisher idea \cite{Fisher:67,Elliott:06} and the reliable estimates of the 
Hills and Dales Model \cite{Bugaev:04b} represents a sum of all surface deformations of a given bag.}
Choosing sufficiently large radius $R$ and very large height $L \gg R$ (thermodynamic limit) one can see 
that the last term in Eq.(\ref{EqV}) vanishes. Clearly, the corrections coming from the Coulomb part in the 
string free energy $F_{str}$ or from different parameterization of the Fisher topological term in 
Eq.(\ref{EqI}) and in Eq.(\ref{EqV}) should also vanish in this limit. {Note that the term proportional 
to $\ln L$ is present in the string free energy $F_{str}$ as well, but it was not analyzed in \cite{Patel} 
due to the same reason.}

The last result shows that for the thin strings $R \rightarrow 0$ (compared to $L$) or for the vanishing 
thermal pressure $p_v \rightarrow 0 $ there exists a simple interrelation between the colour string tension and 
the surface tension of the QG bag $\sigma_{str} (T) \approx  \sigma_{surf} (T)\, 2 \pi R$, but in general 
Eq.(\ref{EqV}) determines the temperature dependence of string radius 
\begin{equation}\label{EqVI}
R^\pm = \frac{\left[\sigma_{surf} (T) \pm \sqrt{\sigma_{surf}^2(T) - \frac{p_v(T)\sigma_{str} (T)}{\pi}}\right]}{p_v (T)}\,,
\end{equation}
if the temperature dependences of $\sigma_{str} (T)$, $\sigma_{surf} (T)$ and $p _v (T)$ are known. On the 
other hand it is also possible to determine the $T$-dependence of the surface tension of bags
\begin{equation}\label{EqVII}
\sigma_{surf} (T) = \frac{\sigma_{str} (T)}{ 2 \pi R} ~ + ~ \frac{1}{2} \, p_v (T) R \,,
\end{equation}
if $R(T)$, $\sigma_{str} (T)$ and $p_v (T)$ are known. In fact, Eq.(\ref{EqVI}) already gives us the 
following radius independent inequality for bag surface tension 
\begin{equation}\label{EqVIII}
\sigma_{surf} (T)^2 ~ \ge ~ \frac{p_v(T) \sigma_{str} (T)}{\pi}      \,,
\end{equation}
which demonstrates that for a confining string, i.e. for $\sigma_{str} (T) > 0$, the bag surface tension can 
vanish only and only for $ p_v(T) \le 0$, i.e. only for negative or zero values of thermal pressure! The 
latter is clearly seen, if one substitutes $ \sigma_{surf} (T) \rightarrow 0 $ into Eq.(\ref{EqVI}) and 
considers $\sigma_{str} (T) > 0$, i.e.
\begin{equation}\label{EqIX}
R^-\biggl|_{\sigma_{surf} (T) \rightarrow 0} ~ \rightarrow ~ \sqrt{\frac{\sigma_{str} (T)}{- \pi\, p_v(T)}}\,,
\end{equation}
which is real for $p_v(T) \le  0$ only. 

Eq.(\ref{EqVII}) allows us to estimate roughly the surface tension at $T=0$. Taking the typical value of the 
bag model pressure as $p_v (T=0) = - (0.25)^4$ GeV$^4$,  $R=0.5$ fm  and $\sigma_{str} (T=0) = 
(0.42)^2$ GeV$^2$ \cite{StrTension2}, one finds from (\ref{EqVII}) that 
$\sigma_{surf} (T=0) = (0.2229~ {\rm GeV})^3 + 0.5\, p_v\, R\approx (0.183~{\rm GeV})^3 \approx 157.4$ MeV fm$^{-2}$.  
Our estimate is larger than the ones which are familiar to the astrophysics community \cite{Ignat:1} but we 
would like to emphasize it is based on conservative parameter values in Eq.(\ref{EqVII}). Optimizing the 
radius $R$ and bag constant value we could get perfectly suitable magnitude of the surface tension at 
vanishing temperature.

%%%%%%%%%%%%%%%%%%%%%%%%%%%%  Sect III

\section{Thermodynamics of Cylindrical QG Bag}

The above results allow us to tune the interrelation with the colour string model and to study the bag 
surface tension near the cross-over to QG plasma phase. The LQCD data 
%%%simulations 
indicate that at large $R$ the 
string tension behaves as \cite{StrTension,ConfString3}
\begin{equation}\label{EqX}
 \sigma_{str}^{LQCD}  ~ \approx ~ \frac{\ln\left( L/L_0 \right)}{R^2 } C \,,
\end{equation}
where $L_0$ and $C$ are some positive constants.  
{Such a behavior in a confined phase can be easily understood within the confining string model 
\cite{Patel}. Indeed, at low $T$  the string energy is proportional to the separation $L$. As the temperature 
increases, the flux tube starts to oscillate and its length exceeds the separation distance $L$. Eventually 
the flux tube travels all over the available space before ending on the colour charges.} 

{In fact, a very similar explanation emerges from the view point of elongated cylinder despite the 
different air of Eq.(\ref{EqX}) and Eq.(\ref{EqV}).}
%%Comparing  (\ref{EqX}) and (\ref{EqV})  one may wonder whether  
%%they are similar to each other.  
Noting that the weak $L$ dependence of the LQCD string tension (\ref{EqX}) still has to be accurately explored 
for the large values of quark-antiquark pair separation $L$, here we, however, would like to study the  
collapsing string tension at fixed $L \gg L_0$ {(or better to say, the string melting)} and the  
temperature approaching the cross-over temperature $T_{tr}$ from below, i.e. for $T \rightarrow T_{tr} - 0$, 
{which according to the LQCD \cite{StringEntropy} and to the flux tube model \cite{Patel} are equal.  
Similar to \cite{Patel} we assume that in the infinite available volume} 
$\sigma_{str} (T)\rightarrow + 0 ~ \Rightarrow~ R \rightarrow \infty$ in such a way that 
($\omega_k = const \sim \ln(L/L_0)$)
\begin{equation}\label{EqXI}
\sigma_{str} (T) \, R^k \rightarrow \omega_k >0 \,,
\end{equation}
thereby extending a range for the power $k>0$ {to study more general case, since} the formal 
expressions for thermodynamic functions are valid {both for positive and negative values of $k$.
The value of constant $\omega_k > 0$ is not of crucial importance  for us here because we are interested in 
the qualitative analysis while it becomes quite essential for the quantitative estimates.}

{The effect of outer baryonic charge in the system of quark-antiquark pairs (with zero net baryonic 
charge) can be accounted for by the dependence of string melting temperature $T_{tr}$ on the baryonic chemical 
potential $\mu_b$. Apparently it leads to the temperature $T_{tr}$ decreasing as a function of $\mu_b$ as 
expected by the QCD phenomenology \cite{Shuryak:08}. Thus, in what follows we also assume that Eq.(\ref{EqXI}) 
is valid for the non-zero values of $\mu_b$. Performing our analysis at fixed values of $\mu_b$ we have no 
need to introduce the particular dependence of all quantities (including $\omega_k$ in Eq.(\ref{EqXI})) on it.  
We have to keep in mind only that the temperature $T_{tr}$ may change with $\mu_b$. 

The surface tension coefficient can be affected by $\mu_b > 0$ as well, but the main result of Eq.(\ref{EqV}) 
remains obviously valid.} Now neglecting the last term on the right hand side of Eq.(\ref{EqV}) for $L \gg R$ 
one can calculate the thermal pressure (treating Eq.(\ref{EqXI})) as
\begin{eqnarray}\label{EqXII}
&&p_v (T)   = {\textstyle  
2\, \frac{ \sigma_{surf} (T)}{R} - \frac{\sigma_{str} (T) }{\pi R^2} }
  \rightarrow  
{\textstyle 
\left[  \frac{\sigma_{str}}{\omega_k} \right]^{\frac{1}{k}}  \left[ 2 \, \sigma_{surf}  ~  - ~ \frac{\omega_k }{\pi}
\left[  \frac{\sigma_{str}}{\omega_k} \right]^{\frac{k+1}{k}} 
\right] 
}  \,. 
\end{eqnarray}
Similarly, neglecting the Fisher topological term in Eq.(\ref{EqIII}) at $L \gg R$ (for a cylindrical bag of 
radius $R$ and height $L \gg R$) one obtains the total bag pressure  
\begin{eqnarray}\label{EqXIII}
&&p_{tot} = p_v (T) {\textstyle - \frac{ \sigma_{surf} (T)}{R} \equiv \frac{ \sigma_{surf} (T)}{R} - \frac{\sigma_{str}}{\pi R^2} }  \rightarrow 
{\textstyle  
\left[ \frac{\sigma_{str}}{\omega_k} \right]^{\frac{1}{k}} \left[ \sigma_{surf} ~ - ~ \frac{\omega_k}{\pi}
\left[ \frac{\sigma_{str}}{\omega_k} \right]^{\frac{k+1}{k}} 
\right]   
} \, ,
\end{eqnarray}
and its  total entropy density 
\begin{eqnarray}\label{EqXIV}
\hspace*{-0.5cm}&&s_{tot} = \frac{\partial ~p_{tot}}{\partial~T} \rightarrow   
{\textstyle 
 \frac{1}{k\,\sigma_{str} } \left[ \frac{\sigma_{str}}{\omega_k} \right]^{\frac{1}{k}}  
\frac{\partial ~\sigma_{str} }{\partial ~T} \, \sigma_{surf} ~ +~ 
\left[  \frac{\sigma_{str}}{\omega_k} \right]^{\frac{1}{k}}  
\frac{\partial~\sigma_{surf} }{\partial ~T} 
~ - ~ \frac{k+2}{\pi\, k} \left[ \frac{\sigma_{str}}{\omega_k} \right]^{\frac{2}{k}} \frac{\partial ~\sigma_{str} }{\partial ~T} 
}  \,. ~~~
\end{eqnarray}
The mechanical stability of the cylindrical bag means an equality of the total bag pressure Eq.(\ref{EqXIII}) to 
the outer pressure, but the thermodynamic stability requires positive value for the entropy density (\ref{EqXIV}).  
To quantify the latter we adopt the following parameterization of the string tension for $T \rightarrow T_{tr} - 0$
\begin{equation}\label{EqXV}
\sigma_{str} (T) = \sigma_{str}^0\, t^\nu, \quad {\rm where} \quad \quad t \equiv \frac{T_{tr}-T}{T_{tr}}\rightarrow +0\,,
\end{equation}
with $\sigma_{str}^0 >0$ and $\nu>0$ (for example, in the simplest case $\nu =1$ at $\mu=0$ \cite{Patel}, 
but the other values can also be valid). Then the total entropy density  of the cylindrical bag becomes 
\begin{eqnarray}\label{EqXVI}
s_{tot}  & \rightarrow   &
{\textstyle 
\left[  \frac{\sigma_{str}^0\, t^\nu}{\omega_k} \right]^{\frac{1}{k}}  ~ \biggl\{- \frac{\nu}{k\, T_{tr} }    
 \,  \frac{ \sigma_{surf}  }{t}~ +~ 
\frac{\partial ~\sigma_{surf} }{\partial~T}  
+
 \left. \frac{(k+2)\nu}{\pi\, k}\left[ \frac{\sigma_{str}^0\,t^\nu}{\omega_k}\right]^{\frac{1}{k}} \frac{\sigma_{str}^0\, t^{\nu-1} }{T_{tr}}  
\right\} 
}  \,. 
\end{eqnarray}
The model of quark gluon bag with surface tension \cite{Bugaev:07,CritPoint} predicts that everywhere at the 
cross-over line, except for the (tri)critical endpoint, the surface tension coefficient $\sigma_{surf}$ is 
non-zero and its derivative $\frac{\partial ~\sigma_{surf}}{\partial ~T}$ is finite at $t \rightarrow +0$. 
Remembering this requirement one finds from Eq.(\ref{EqXVI}) that its first term of the right hand side 
dominates and, hence, we receive  
\begin{equation}\label{EqXVII}
s_{tot} \rightarrow - \left[  \frac{\sigma_{str}^0\, }{\omega_k} \right]^{\frac{1}{k}}  ~ \frac{\nu}{k\, T_{tr} } 
 \,   \sigma_{surf} (T_{tr})\,\,  t^{ \frac{\nu}{k} - 1}  > 0  \,,
\end{equation}
which means that at $T \rightarrow T_{tr} - 0$ the surface tension coefficient must be negative 
$\sigma_{surf} (T_{tr}) < 0$. Actually, this result brings nothing surprising since {the calculations 
of surface partitions for physical clusters \cite{Bugaev:04b} and} the model of quark gluon bag with surface 
tension with tricritical \cite{Bugaev:07} and critical endpoints \cite{CritPoint} predict also that at low 
baryonic densities the deconfining phase transition degenerates in a cross-over just because the surface tension 
coefficient of large bags becomes negative in this region {(for more details see next section).}
Eq.(\ref{EqXVII}) clearly shows that  the colour string model shares the possibility of negative values of bag surface 
tension coefficient available in the cross-over region. 

However, considering the above results in the context of the LQCD data we should mention it is unlikely
that the current calculations on the finite lattices allow us to claim an immutable validity of Eq.(\ref{EqXI}). 
Indeed, an analysis of the quarkonium spectra in the deconfined phase teaches they are surviving (not melted) far 
behind the critical temperature signaling that the string tension does not completely vanish at high temperatures 
\cite{StringEntropy}. Apparently, it implies the modification of Eq.(\ref{EqXI}) and new phenomenological inputs 
inevitable.
The LQCD data for the free energy (and entropy) of the colour string \cite{StringEntropy} demonstrate an extremely  
fast increase at approaching the cross-over temperature from below 
{although the mean free energy of the string $\langle F_{str} \rangle$ is finite at $L \rightarrow \infty$.
In order to make these LQCD results instrumental and not to modify Eq.(\ref{EqXI}) at the scale of finite lattice 
of the spatial size $R_{lat}$ the mean free energy of the string at $L \rightarrow \infty$ can be subdivided in two 
parts. One part with the probability $W(L)$ corresponds to the strings of infinite values of 
$F^\infty_{str} = \sigma_{str} L$ in the limit $L \rightarrow \infty$ and another part for the strings having the finite values of free energy 
$F^0_{str}$, i.e. $\langle F_{str}\rangle = F^\infty_{str} W + F^0_{str}(1 - W)$. Then it seems reasonable to assume 
that the entropy of the strings of finite free energy is finite and smooth function of $t\rightarrow +0$. Now to 
explain the behavior of the lattice free energy and entropy  of the colour string basing on Eqs.(\ref{EqXI}) and 
(\ref{EqXV}) we have to expect $W(L)$ behaving as $W \sim [L \ln (L/L_0)]^{-1}$ at $L \rightarrow \infty$ and 
fixed $R$. If it is well argued to use the above results for $R < R_{lat}$ and $L \rightarrow \infty$ to compare 
them with the LQCD data, then using the thermodynamic identity $\frac{\partial F^\infty_{str}}{\partial T} = 
- S^\infty_{str}$, Eq.(\ref{EqV}) and the first equality in  Eq.(\ref{EqXIII}) one obtains the lattice entropy $S^\infty_{str}$ 
%%%of the first part 
at fixed $L$ as}
\begin{equation}\label{EqAI}
S^\infty_{str} ~= ~ - ~ \frac{\frac{\partial\sigma_{surf}}{\partial~ T}\pi R - s_{tot}\pi R^2}{1+\frac{\sigma_{surf}\pi R}{k\,\sigma_{str}} - 
\frac{2\, p_{tot}\pi R^2}{k\,\sigma_{str}}} L \,.
\end{equation}

The colour  tube radius for vanishing $\sigma_{str}$ can be found from Eq.(\ref{EqVI}) as 
$R\rightarrow \frac{2\,\sigma_{surf}}{p_v}$. It helps to simplify the denominator in Eq.(\ref{EqAI}) and then 
neglecting the term with the derivative of surface tension in the numerator of Eq.(\ref{EqAI}) (the term with 
$s_{tot}$ is larger) one finally receives 
\begin{equation}\label{EqAII}
S^\infty_{str} ~ \rightarrow ~ - ~ \frac{s_{tot}\, k\, \sigma_{str} R}{\sigma_{surf}} L ~ = ~ - ~\frac{s_{tot}\, k\, \omega_k    }{ \sigma_{surf}  R^{k-1} } L
 \,,
\end{equation}
where at the last step of deriving (\ref{EqAII}) we used Eq.(\ref{EqXI}).

It is easy to verify (using Eqs.(\ref{EqXV}) and (\ref{EqXVII})) that despite the approximations done Eq.(\ref{EqAII})
is nothing more than $(-\frac{\partial \sigma_{str}}{\partial ~T}\, L)$.
Now it becomes clear that  the contribution of this  term into the total entropy is finite  
$S^\infty_{str} W  < \infty $ for $L\rightarrow \infty$ and $R < R_{lat}$. Again we see that  it should be $\sigma_{surf} (T_{tr}) < 0$
to provide the positive values of  $S^\infty_{str}$ and  $s_{tot}$.

Besides, it is clear from (\ref{EqAII}) that with $t$ decreasing to a minimal value and $R$ approaching 
$R_{lat}$ 
one has $S^\infty_{str} W \sim t^{\nu-1}$, i.e. the entropy part of the strings of infinite free energy increases 
quickly for $\nu <1$. A formal extension of this result to $R \rightarrow \infty$ leads to a divergency of 
$S^\infty_{str} W \sim t^{\nu-1}$ at $t=0$. However, to study the behavior of the lattice entropy with more accuracy 
it is necessary to assume a certain behavior of the string tension in the vicinity of $t =0$.

In order to demonstrate the possibility for entropy density Eq.(\ref{EqXIV}) to be divergent at $t = 0$ even for 
the lattice of finite size we consider the simplest modification of Eq.(\ref{EqXV}) which clarifies the fact of 
string tension survival at $t = 0$. It has still a finite magnitude but is going down for $T > T_{tr}$ 
\cite{StringEntropy}. Actually, such a modification accounts for that the radius of the colour string $R$ in 
(\ref{EqXI}) can not exceed the lattice size. We choose the simplest parameterization 
\begin{eqnarray}\label{EqXVIII}
\sigma_{str} (T) =  
\left\{ \begin{array}{rr}
\sigma_{str}^{tr}  + \sigma^- \cdot t^{\nu^-}   \,,  &\hspace*{0.1cm}  T \rightarrow T_{tr}  - 0 \,,\\
%& \\
\sigma_{str}^{tr}  + \sigma^+ \cdot (-t)^{\nu^+}   \,,  &\hspace*{0.1cm}  T \rightarrow T_{tr}  + 0 \,,
\end{array} \right. 
\end{eqnarray}
where $t \equiv (T_{tr}-T)/T_{tr} $, the string tension coefficient at the cross-over temperature 
$\sigma_{str}^{tr} > 0$ is small but non-zero, $ \sigma^\pm$ and $\nu^\pm > 0$ are some non-zero  constants. Now 
using Eq.(\ref{EqXVIII}) we find the string tension derivative 
\begin{eqnarray}\label{EqXIX}
\frac{\partial \sigma_{str} }{\partial~ T}  =  \frac{1}{T_{tr}}
\left\{ \begin{array}{rr}
- \sigma^- \, \nu^- \, t^{\nu^- - 1}  < 0 \,,  &\hspace*{-0.1cm}  T \rightarrow T_{tr}  - 0 \,,\\
%& \\
\sigma^+ \, \nu^+ \, (-t)^{\nu^+ -1} < 0  \,,  &\hspace*{-0.1cm}  T \rightarrow T_{tr}  + 0 \,. 
\end{array} \right. 
\end{eqnarray}
Since the LQCD data show that the string tension coefficient is monotonically decreasing with $T$ on both sides of 
$T= T_{tr} $, whereas the entropy increase (decrease) is very fast for 
$T \rightarrow T_{tr} - 0$ ($T \rightarrow T_{tr}  + 0$) \cite{StringEntropy}, we conclude from Eq.(\ref{EqXIV}) 
and Eq.(\ref{EqXIX}) that such a behavior can be provided by $- \frac{\partial\sigma_{str}}{\partial ~T}\, L\, W(L)$ 
if the following inequalities are valid $\sigma^->0$, $\sigma^+ < 0$, $\sigma_{surf} (T_{tr}) < 0$ and 
$\nu^\pm < 1$. In proving this statement one has to account the finiteness of $\sigma_{surf} (T_{tr})$ and its 
derivative and the smallness of $\sigma_{str}^{tr}$ compared to  $|\sigma_{surf} (T_{tr})|$.

Thus, the presence of a non-zero surface tension in the entropy density (\ref{EqXIV}), (\ref{EqXVI}) and 
(\ref{EqXVII}) naturally explains also the origin of extremely fast entropy increase close to the cross-over 
temperature for any lattice sizes. This effect was called `mysterious' in Ref.\cite{Shuryak:08} but we note
here that a similar behavior should be inherent in the energy density as well since the pressure Eq.(\ref{EqIII}) 
is finite. 
{In principle, one could choose more sophisticated T-dependence of the string tension $\sigma_{str}$ 
in order to describe the finite maximum of the lattice entropy but such a task requires, first of all, the LQCD
data of very high quality to be conclusive. 
}

Furthermore, the equations (\ref{EqXIV}), (\ref{EqXVI})--(\ref{EqXIX}) derived together with Eq.(\ref{EqXI}) 
provide us now with a principal possibility to determine the QG bag surface tension along the cross-over line in 
the $(\mu_b - T)$ - plane directly from the LQCD data. Such results are of vital importance for quantitative 
estimates while dealing with the model of quark gluon bags with surface tension \cite{Bugaev:07,CritPoint} and 
its generalizations \cite{FWM,Reggeons:08,Bugaev:08}.

%%%%%%%%%%%%%%%%%%%%%%%%%%%%  Sect IV

\section{The (Tri)critical Endpoint Vicinity}

Another special and interesting possibility to be considered is if the surface tension coefficient in 
Eq.(\ref{EqXVI}) vanishes simultaneously with the string tension as 
\begin{eqnarray}\label{EqXX}
\sigma_{surf} (T) =  
\left\{ \begin{array}{rr}
\sigma^-_{surf}  \cdot t^{\zeta^- }   \,,  &\hspace*{0.1cm}  T \rightarrow T_{tr}  - 0 \,,\\
%& \\
\sigma^+_{surf}  \cdot (-t)^{\zeta^+ }   \,,  &\hspace*{0.1cm}  T \rightarrow T_{tr}  + 0 \,,
\end{array} \right. 
\end{eqnarray}
or, in other words, the surface tension vanishes at the cross-over line. As shown in \cite{Bugaev:07,CritPoint}  
such a situation takes place at the tricritical or critical endpoint only, whereas in the example of 
Eq.(\ref{EqXVII}) the surface tension coefficient is considered at the values of baryonic chemical potential 
which are smaller than that of the (tri)critical endpoint. Here we use the surface tension in the form introduced 
in \cite{CritPoint} intending to apply it to the analysis of (tri)critical endpoint. 

Choosing this simple surface tension parameterization, Eq.(\ref{EqXX}), we follow the original Fisher idea 
\cite{Fisher:67} to explain the temperature dependence of surface free energy for $\zeta^\pm =1$. According to 
that idea the surface free energy consists of two terms, the surface energy of a bag of volume $V$ as 
$\sigma^\pm_{surf} V^\frac{2}{3}$ and the term $- T \sigma^\pm_{surf} T_{tr}^{-1} V^\frac{2}{3}$ which comes 
from the surface entropy $\sigma^\pm_{surf} T_{tr}^{-1} V^\frac{2}{3}$ in the original model \cite{Fisher:67}.
Note that the surface entropy of a bag of volume $V$ counts its degeneracy factor or the number of ways to have 
such a bag with all possible surfaces. This result was generalized to $\zeta^\pm > 1$  using the exact solutions 
of the Hills and Dales Model \cite{Bugaev:04b} while we consider the interval $0 < \zeta^\pm < 1$. The Hills and 
Dales Model naturally explains the negative values of the surface free energy. They are originated by the 
dominance of bags of complicated (non-spherical) shapes whose number of states (exponential of entropy) simply  
exceeds their suppression by the Boltzmann factor.  

Substituting (\ref{EqXX}) into (\ref{EqXVI}) and using (\ref{EqXV}) we get for $T \rightarrow T_{tr} -0$ in
thermodynamic limit
\begin{eqnarray}\label{EqXXI}
&&s_{tot} \rightarrow  \textstyle \left[\frac{\sigma_{str}^0\, t^\nu}{\omega_k} \right]^{\frac{1}{k}} ~\left\{- \left[ \zeta^- +\frac{\nu}{k } \right]
 \,  \frac{ \sigma_{surf}^-  t^{\zeta^- -1}}{T_{tr} }
 +  
    \frac{(k+2)\nu}{\pi\, k} \left[  \frac{\sigma_{str}^0\,t^\nu}{\omega_k} \right]^{\frac{1}{k}}  \frac{\sigma_{str}^0\, t^{\nu-1} }{T_{tr}}  
\right\}
 \,. 
\end{eqnarray}
In contrast to (\ref{EqXVII}) the surface tension term in (\ref{EqXXI}) dominates for $\zeta^- < \frac{k+1}{k}\nu$ 
only and it implies $\sigma^-_{surf} <0$. Let us note the latter contradicts to critical endpoint existence 
conditions found in \cite{CritPoint} but such an inequality does not affect the tricritical endpoint existence 
\cite{Bugaev:07, FWM}. The entropy density would diverge in this case for $\zeta^- < 1 - \frac{\nu}{k}$ (which 
leads to $\nu < k$) whereas for $1 - \frac{\nu}{k}\le \zeta^- < \frac{k+1}{k} \nu$ (which is true for 
$\nu \ge \frac{k}{k+2}$) the entropy density would vanish at the tricritical endpoint.  
If, however, $\zeta^- > \frac{k+1}{k} \nu$ then the last term on the right hand side of Eq.(\ref{EqXXI}) dominates 
and, hence, the entropy density is obviously positive for $k >0$ and $\nu>0$. Under this condition for $\zeta^-$ 
both the critical and tricritical endpoints may exist and the entropy density diverges at $t = 0$ for 
$\nu < \frac{k}{k+2}$. 

Considering now Eq.(\ref{EqXVIII}) as the finite size analog of (\ref{EqXV}) we have for the entropy density in the 
vicinity of $ \mp t \rightarrow + 0$   
\begin{eqnarray}\label{EqXXII}
s_{tot} &\rightarrow & \textstyle \left[  \frac{\sigma_{str}^{tr} }{\omega_k} \right]^{\frac{1}{k}} 
\frac{ \left(\pm \sigma^\pm \right)}{T_{tr} }\,  \nu^\pm    \left(  \mp t  \right)^{\nu^\pm - 1}
 \, \left\{ 
   \frac{  (\mp t)^{\zeta^\pm } \, \sigma_{surf}^\pm}{\sigma_{str}^{tr} }
 -   \frac{(k+2)}{\pi\, k} \left[\frac{\sigma_{str}^{tr}}{\omega_k} \right]^{\frac{1}{k}} 
\right\}
  \nonumber \\
&  + &  \textstyle
\left[ \frac{\sigma_{str}^{tr} }{\omega_k} \right]^{\frac{1}{k}} \frac{\sigma_{surf}^\pm}{T_{tr}}\, \zeta^\pm \, (\mp t)^{\zeta^\pm -1 } 
 \,. 
\end{eqnarray}
This equation has more complicated structure being compared to Eq.(\ref{EqXXI}). For $\nu^\pm > 0$ and $\zeta^\pm > 0$, 
however, the term which results from the surface tension in (\ref{EqXXII}) (proportional to 
$(\mp t)^{\nu^\pm + \zeta^\pm -1} $) can not dominate anymore. Nevertheless, the term originated by its derivative 
(the last term in (\ref{EqXXII})) is dominating for $\nu^\pm > \zeta^\pm$ implying $\pm \sigma_{surf}^\pm > 0$, 
since according to (\ref{EqXIX}) $\pm \sigma^\pm < 0$. And as before this possibility, $\pm \sigma_{surf}^\pm > 0$, 
is inconsistent with the critical endpoint existence \cite{CritPoint}. For $\nu^\pm < \zeta^\pm$ the second term on 
the right hand side of Eq.(\ref{EqXXII}) dominates and it leads to the positive value of entropy density (\ref{EqXXII}). 

Clearly, the entropy density Eq.(\ref{EqXXII}) would diverge at $t=0$ for either $\nu^\pm < 1$ or $\zeta^\pm < 1$
being essentially different from what we found out at analyzing (\ref{EqXXI}). The important message of this 
comparison is that one has to be very careful at operating with the LQCD results to extract a physical information 
above a cross-over on the phase diagram.

%%%%%%%%%%%%%
%%%KKK

%%%KKK 

\section{Conclusions}
{
Here we develop the novel approach to the bag phenomenology of deconfinement using quark gluon bag model with 
surface tension \cite{Bugaev:07, FWM,Reggeons:08,Bugaev:08, CritPoint} as an effective tool for exploring the
particular region of phase diagram of strongly interacting matter. Analyzing the free energies of cylindrical 
bags and confining colour strings we find out the quite general alignment connecting the string tension with 
bag surface tension and thermal pressure. This relation makes it possible, in principle, to estimate the bag 
surface tension directly from the LQCD measurements. 
We use the relation derived to study (under the plausible assumptions) the string entropy behavior 
in the cross-over vicinity, i.e. for (almost) vanishing string tension, taking into account the finite lattice 
size in the LQCD simulations. We find out the divergent behavior of the entropy density for the finite values 
of bag surface tension at the cross-over. The relation (obtained in Eq.(\ref{EqAII}) between the cylinder bag 
entropy density and the lattice entropy $S^\infty_{str} W(L)$ allows us to naturally explain the `mysterious' 
\cite{Shuryak:08} maximum of $S^\infty_{str} W(L)$ measured in the LQCD simulations at the cross-over. Moreover,  
drawing the corresponding LQCD results we determine that the surface tension coefficient is amazingly negative
at the cross-over in conformity with the prediction of quark gluon bag model with surface tension 
\cite{Bugaev:07, FWM,Reggeons:08,Bugaev:08, CritPoint}. This model says that the only physical reason for 
transforming the first order deconfining  phase transition into the cross-over at low baryonic chemical 
potentials is just the negative values of the surface tension coefficient of bags in this region of phase
diagram. Apparently, the LQCD data give us the quite serious confirmation of this finding.  Comparing
the behavior of entropy density in the vicinity of (tri)critical endpoint for infinite and finite systems we
find out the different quantitative results although their qualitative behaviors are similar. Besides,  the 
estimates obtained could be much more conclusive if rely upon the reliable LQCD data at non-zero baryonic
densities. Unfortunately, we are still only at the starting point on this road and due to this fact the bag 
model with surface tension looks more suitable for phenomenological analysis, at least, today.} 

The  formulae derived allow us to estimate the bag surface tension at zero temperature drawing the standard  
values of thermal pressure as given by the equation of state of bag model. Surprisingly it occurs rather large  
$\sigma_{surf} (T=0) \approx (0.183~{\rm GeV})^3 \approx 157.4$ MeV fm$^{-2}$ if compared to the estimates 
received in the other phenomenological models. Surely, more accurate LQCD analysis of the bag surface tension 
will be done in a time to come but today we believe our estimate based on the conservative values of parameters 
provides, in a sense, its lower bound which could be practical for studying the early Universe. In one of those
scenarios \cite{NoBoiling} the quark matter may survive the 'boiling' period of the early Universe if  
$\sigma_{surf} > (0.178~{\rm GeV})^3 $. However, the idea of quark matter survival to the present days discussed 
in \cite{Witten} requires a revision in view of the negative values of surface tension coefficient at the 
cross-over. Moreover, it definitely needs an additional phenomenological input since the large QG bags have very
short life-time as shown within the models like \cite{FWM,Reggeons:08,Bugaev:08} or within the Hagedorn-Mott 
resonance gas model \cite{Blaschke:03}.

{\bf Acknowledgments.} We are thankful to S.V. Molodtsov and V.K. Petrov for fruitful discussions and pointing 
out some LQCD results. The research was supported in part by the Program 'Fundamental Properties of Physical 
Systems under Extreme Conditions' launched by the Section of Physics and Astronomy of National Academy of 
Sciences of Ukraine. K.A.B. acknowledges the partial support of  the Fundamental Research State Fund of Ukraine, 
Agreement No F28/335-2009 for the Bilateral project FRSF (Ukraine) -- RFBR (Russia).

%\acknowledgments

\end{document}